# Validation of critical maneuvers based on shared control

Mauricio Marcano, Joseba Sarabia, Asier Zubizarreta and Sergio Díaz

*Abstract*— This paper presents the validation of shared control strategies for critical maneuvers in automated driving systems. Shared control involves collaboration between the driver and automation, allowing both parties to actively engage and cooperate at different levels of the driving task. The involvement of the driver adds complexity to the control loop, necessitating comprehensive validation methodologies. The proposed approach focuses on two critical maneuvers: overtaking in low visibility scenarios and lateral evasive actions. A modular architecture with an arbitration module and shared control algorithms is implemented, primarily focusing on the lateral control of the vehicle. The validation is conducted using a dynamic simulator, involving 8 real drivers interacting with a virtual environment. The results demonstrate improved safety and user acceptance, indicating the effectiveness of the shared control strategies in comparison with no shared-control support. Future work involves implementing shared control in drive-by-wire systems to enhance safety and driver comfort during critical maneuvers. Overall, this research contributes to the development and validation of shared control approaches in automated driving systems.

## I. INTRODUCTION

The field of shared control has made significant progress in recent years, exploring various user-oriented strategies [1]–[4]. The concept of shared control involves collaboration between the driver and automation, where both parties are actively engaged and cooperate at different levels of the driving task. In shared control strategies, both the automated system and the driver can interact with the vehicle. Additionally, it focuses on allowing a smooth interaction between the two and can assign full control to either the human or the automation in safety-critical scenarios.

In this sense, shared control approaches add an additional layer of complexity compared to other automated driving features. This is due to the involvement of the driver in the control loop of the vehicle. In fact, driver interaction is a key issue in this type of control, which must also be considered during the validation process.

As the field of shared control continues to expand, the need for robust validation methodologies becomes increasingly important. In particular, validation needs to consider not only the performance evaluation of the automated system assistance, but also the driver's acceptance of shared control and the interface designed for it. In addition to analyzing the performance on standardized maneuvers in a simulated environment, questionnaires and studies involving real drivers have been proposed in the literature for this purpose [5], [6].

Validating automated driving systems is essential before implementing them in actual road vehicles. Many studies have concentrated on developing new features related to vehicle control, but the growing complexity of these functionalities has underscored the need for new validation procedures. Traditionally, driving system features were validated through track testing. However, the increasing complexity and various potential scenarios associated with new features have led to a significant rise in the time and cost required for track testing. Therefore, alternative approaches utilizing vehicle simulations have been suggested to validate new automated driving systems, aiming to address uncertainties arising from perception and localization systems [7], [8].

In line with this, the AUTOEV@L project seeks to establish a modular architecture for automated driving features, including shared control approaches, along with their corresponding verification and technological validation.

In this paper, the aim is to showcase the validation of two critical maneuvers using algorithms based on shared control. The first maneuver focuses on overtaking in low visibility situations, utilizing the vehicle's automation improved reaction times to assist the driver in executing overtaking maneuvers. The second maneuver evaluates lateral evasive actions, where the shared control system promptly and responsively supports the driver in avoiding collisions. The proposed shared control approach primarily focuses on the lateral control of the vehicle, specifically the steering wheel, with an arbitration module determining the level of automation assistance provided to the driver, which represents the steering wheel torque intensity.

To ensure comprehensive user validation of such systems, a dynamic simulator has been used, in which real drivers interact with a virtual environment. A study with 8 drivers has been conducted to validate the proposed maneuvers. Using this testing platform provides a realistic and immersive environment, facilitating a seamless transition to subsequent validation in real-world vehicle settings, while also ensuring the proposed maneuvers' repeatability and test driver safety.

The paper's rest is structured as follows: Section II details the proposed modular architecture for automated vehicles with shared control functionalities. Section III details the shared control algorithm used for lateral control. Section IV details the design of the arbitration system for the two critical maneuvers. Section V provides the verification results for both maneuvers, while Section VI details the results of the study carried out in an immersive driving simulator with 8 drivers. Finally, the paper's most important ideas are summarized in Section VII.

## II. MODULAR ARCHITECTURE

The concept of shared control is a concept that promotes collaboration in real-time between the driver and the automated vehicle. This collaboration is specifically geared towards decision-making and vehicle control. Shared control considers that the driver is in the control loop, and thus, its primary responsibility is to drive. However, it also considers that the automated system may intervene the driver if necessary. The interaction between both agents (driver and the automated system) is carried out by the steering wheel, which is the control mechanism in this shared control system. Hence, in order to define who has the priority in certain scenarios, an arbitration system is needed, which is the decision-making system associated with the shared controller.

As a part of the AUTOEV@L project, a modular architecture for driver-automation interaction is proposed. Figure 1 shows the modules related to shared control and their relationship to the general architecture for automated driving. Notably, shared control requires the inclusion of three additional modules: the arbitration module, the shared control strategy module, and the low-level actuation module. These modules are essential components that enable shared control to work effectively. This way, the shared control approach focuses on ensuring a smooth interaction between the automated system and the driver when using steering wheel to control the lateral position of the vehicle. The proposed architecture is skillfully organized into three levels of cooperation: tactical, operational, and execution [9], as seen in Figure 2.

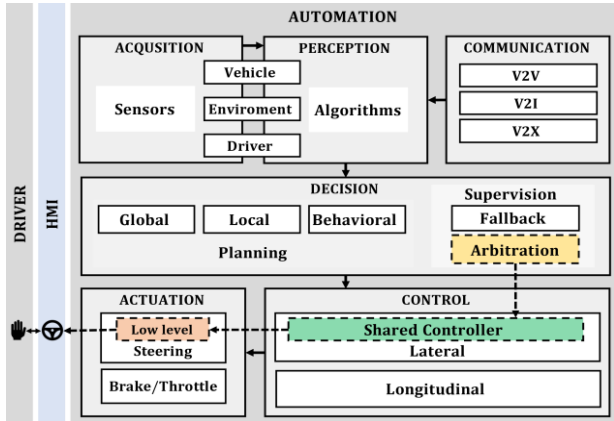

Figure 1. Incorporating modules for shared control into the overall architecture for automated driving.

At the tactical level an Arbitration system is implemented, which decides whether the controller should provide from none to some assistance to the driver or override the driver. For this purpose, fuzzy logic or predefined rules are used to consider inputs from the environment, driver and vehicle to determine intuitive and rational output based on driving and safety requirements.

The operational level, consists of the shared control algorithm, focused on implementing an automated lane centering (ALC) approach, which controls the lateral relative position of the vehicle. For this purpose, a Nonlinear Model-based Predictive Control (NMPC) approach is used, in which different control objectives such as safety, comfort and user interaction; and system constraints (such as maximum torque on the steering wheel) are implemented to obtain the optimum steering wheel torque reference.

Finally, the execution level consists of a PID-based torque control focused on following the optimal reference calculated by the NMPC controller and applying it to the steering wheel by the use of an electric motor. The requirement of such module is that the steering hardware is able to provide an assistance greater than 10 Nm to override the driver intention, control between 1 and 10 ms, and having a variable damping capability to include the stability criteria that will be defined in Section III.

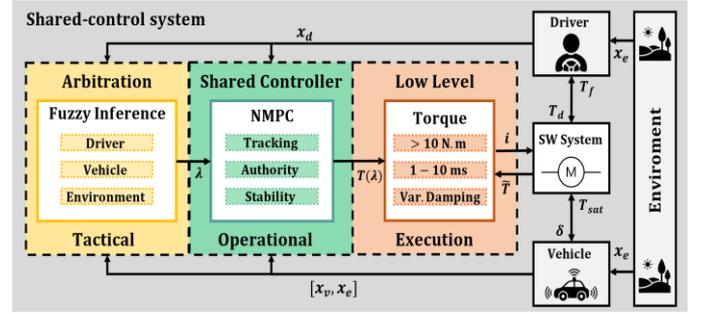

Figure 2. A system of shared control has been detailed, where levels of cooperation are clearly defined.

### III. SHARED-CONTROLLER

As stated in the previous section, the Operational Level of the proposed approach consists of a path tracking lateral controller that computes the optimal torque to be applied to the steering wheel to enable driver interaction. The shared control is based on a automated lane centering (ALC) control, which will try to maintain the vehicle at the center of the lane, but whose assistance will be adapted depending on the command of the Tactical Level, this is, the arbitration system, which will be analyzed in the following section.

The implementation of the lateral controller is carried out by a Nonlinear Model-based Predictive Control (NMPC) approach, in which a nonlinear vehicle model based on the bicycle model (also known as the single-track model) is used (see Figure 3) and equations (1)-(6). Equations (7)-(8) is the road-vehicle model representation in terms of lateral and angular error with respect to the reference trajectory. Equations (9)-(10) represents the steering system model.

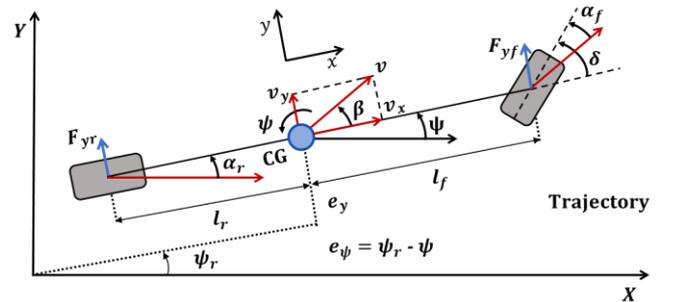

Figure 3. Non-linear vehicle model representation

$$\dot{X} = v_x cos(\Psi) - v_y sin(\Psi) \quad (1)$$

$$\dot{Y} = v_x sin(\Psi) + v_y cos(\Psi) \quad (2)$$

$$\dot{\Psi} = \psi \quad (3)$$

$$\dot{v}_x = (ma_x - F_{yf} sin(\delta) + mv_y \psi)/m \quad (4)$$

$$\dot{v}_y = (F_{yr} + F_{yf} cos(\delta) - mv_x \psi)/m \quad (5)$$

$$\dot{\psi} = (l_f F_{yf} cos(\delta) - l_r F_{yr})/I_z \quad (6)$$

$$\dot{e}_y = v_x sin(e_\Psi) + v_y cos(e_\Psi) \quad (7)$$

$$\dot{e}_\Psi = \dot{\Psi} - \rho v_x \quad (8)$$

$$\dot{\theta} = w \quad (9)$$

$$\dot{w} = (-1/J)(b_\lambda w + T_{sat} - T_{mpc}) \quad (10)$$

$X$ and $Y$ are the absolute coordinates of the center of gravity of the vehicle and $\psi$ is the heading angle, $v_x$ is the longitudinal speed of the vehicle, $v_y$ is the lateral speed, $a_x$ and $a_y$ are the longitudinal and lateral accelerations.

$$F_{yf} = C_{\alpha f}\,(\delta - (v_y + l_f\psi)/v_x) \quad (11)$$

$$F_{yr} = C_{\alpha r}(v_y - l_r\psi)/v_x \quad (12)$$

Equations (11)-(12) represent he front and rear wheels lateral force, $\delta = \theta k_r$ is the steering angle, as a proportional value to the steering wheel angle, $\omega$ is the angular speed of the steering wheel, $m$ is the vehicle mass, $I_z$ is the inertia of the vehicle, $l_r$ and $l_f$ the distances from the center of gravity to the rear and front axles, $J$ is the inertia of the steering wheel, $b_\lambda$ the steering damping, $\rho$ depends on the road curvature, $T_{sat}$ the self-aligning torque and $T_{mpc}$ the assistive torque to be calculated by the MPC algorithm. Considering this model, the following optimization problem is defined for the NMPC.

$$\boldsymbol{x}(k+1) = \boldsymbol{f}(\boldsymbol{x}(k), \boldsymbol{u}(k), \boldsymbol{l}(k)) \quad (13)$$

$$J(\boldsymbol{x}) = \sum_{k=0}^{N-1} \|\bar{\boldsymbol{x}}(k) - \boldsymbol{x}_r(k)\|^2_{W_x} \quad (14)$$

$$J(\boldsymbol{u}) = \sum_{k=0}^{N-1} \|\boldsymbol{u}(k)\|^2_{W_u} + \|\Delta\boldsymbol{u}(k)\|^2_{W_{\Delta u}} \quad (15)$$

$$\min_u \bigl(J(\boldsymbol{x}) + J(\boldsymbol{u})\bigr) \quad (16)$$

$$\boldsymbol{x}_{min} \le \tilde{\boldsymbol{x}} \le \boldsymbol{x}_{max} \quad (17)$$

$$\boldsymbol{u}_{min} \le \tilde{\boldsymbol{u}} \le \boldsymbol{u}_{max} \quad (18)$$

$$\Delta\boldsymbol{u}_{min} \le \Delta\tilde{\boldsymbol{u}} \le \Delta\boldsymbol{u}_{max} \quad (19)$$

Where (13) defines the state-space equation used for prediction, which implements the aforementioned model, and $J(x), J(u)$ are the components of the cost function (14)-(16) to be optimized during the horizon $N$. The state optimization vector is $\bar{x} = [X\ Y\ \Psi\ \psi]$, with $\bar{x} \subseteq x$, $x_r$ is the reference to be followed by the controller, $u = T_{mpc}$ is the optimal torque calculated and $\Delta u = \Delta T_{mpc}$ the time-variation of the calculated torque. $W_x, W_{\Delta u}$ and $W_u$ are the weighting matrices for the reference following, torque variation and torque, respectively. The last three equations of the system (17)-(19) are related to the constraints imposed for the NMPC controller.

In addition, the steering system model incorporates the level of haptic authority ($\lambda$) to help the driver by assigning different levels of stiffness to the torque controller. This is accomplished by utilizing the torque derivative equation $\dot{T}_{mpc} = \lambda\,\Delta T_{mpc}$. Moreover, the authority is related to the maximum steering torque through the torque constraint $|T_{mpc}| < \lambda$. This authority will be modified by the decision-making module depending on the use case, as detailed in the next section.

Moreover, as shown in previous work [10], increasing the authority of the controller makes the system unstable, and therefore, a variable damping value is applied to keep the stability of the vehicle controller $b_\lambda = b\sqrt{(\lambda + 1)/2}$. This damping is introduced in the hardware configuration of the steering wheel motor.

## IV. Decision-Making and Use-Cases

Two use cases considering critical emergency lateral maneuvers will be considered for the implementation of shared based control and its validation: 1) corrective maneuver and 2) evasive maneuver. The main objective is to ensure safety during driving while also providing a comfortable driving experience during these maneuvers, focusing on the interaction and collaboration between the driver and automation through the vehicle's steering system.

As previously detailed, the implementation of shared control approaches requires a decision-making system at the tactical level which is commonly referred to as the arbitration module. Its function is to distribute control authority between the driver and the automation by modifying the weights of the NMPC lateral controller. The design of this system is explained below for the two maneuvers studied within AUTOEV@L, as it is specific to the functionality of the system for each use case.

### A. Corrective maneuver

The corrective maneuver is studied in the context of overtaking on two-way roads. As shown in Figure 4, the vehicle initially operates in assistance mode (1), with adaptive cruise control (ACC) and automated lane centering (ALC) activated. In this scenario, the driver attempts to overtake. If the maneuver is deemed unsafe (2), the system intervenes by increasing the lateral control authority of the vehicle and torque intensity on the steering wheel. On the other hand, if the overtaking is safe, a transition is made from assistance mode to manual mode (3) to allow the driver to perform the maneuver by reducing the assistance torque. Once completed, the assistance mode is automatically reactivated (4).

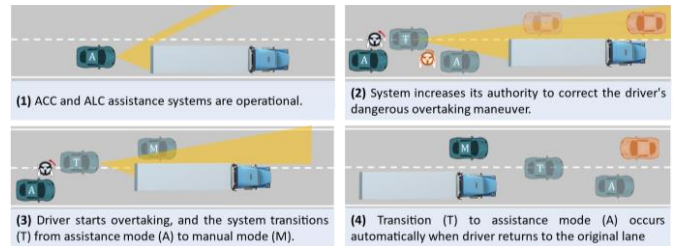

Figure 4. Sequence of scenarios for the study of corrective maneuvers

This behavior is integrated into the decision-making system by means of fuzzy logic techniques. For this purpose, three inputs and one output are defined, which are represented in Figure 5, related by the set of rules in Figure 6, and described below.

- **Input #1 - Vehicle Position:** Represented as the lateral error of the vehicle with respect to the center of the right lane ( ). The membership function labels ([Right - Edge - Left]) represent the different positions of the vehicle on a two-lane road.

- **Input #2 – Driver Intention:** Represented as the derivative of the lateral error of the vehicle (). The membership function labels ([Drift Away - Stay - Return]) represent the driver's intention to leave the lane, stay in the same direction, or return to the lane. This intention is combined with the lateral error to obtain an estimate of the lane-change intention.

- **Input #3 - Maneuver Risk:** Represented as the distance to collision (DTC) between the vehicle and the following vehicle in the left lane. The membership function labels ([Far - Near]) represent the relative distance between the two vehicles, indicating low and high collision risk, respectively.

- **Output #1 - Authority Level**: Represented as the maximum torque in Nm for steering correction (λ), which will be handled to the NMPC controller as a parameter. The membership function labels ([Low – Medium - High]) represent the full range of automated steering assistance, from none to gentle corrections, up to the maximum assistance which, for this use case, is a torque of 8 Nm.

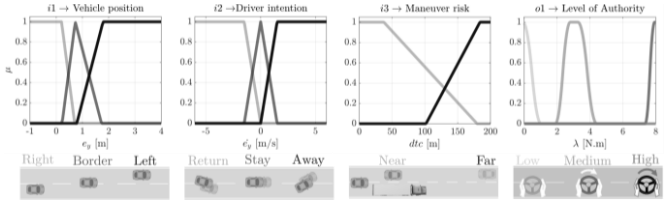

Figure 5. Inputs and outputs of the fuzzy logic-based arbitration system of the correction maneuver

Figure 6. IF-THEN rules of the fuzzy inference system for the arbitration system

### B. Evasive maneuver

The study of evasive maneuvering involves a lane invasion scenario by a motorcycle on a two-way road. In this case, the decision-making system is not based on fuzzy logic, but on a set of rules that modify the configuration parameters of the MPC controller.

The sequence of the maneuver can be observed in Figure 7. At the beginning, the vehicle is in assistance mode (1), which means that the ACC and ALC systems are activated. Upon detecting a lane invasion, the vehicle's displacement limits are modified based on its prediction horizon $N$. These limits can be modified in real-time in the NMPC configuration, which estimates a new prediction that meets these restrictions. At the same time, greater authority is assigned to the controller (2) and (3) until the risk of the maneuver decreases, and the system returns to its initial state (4).

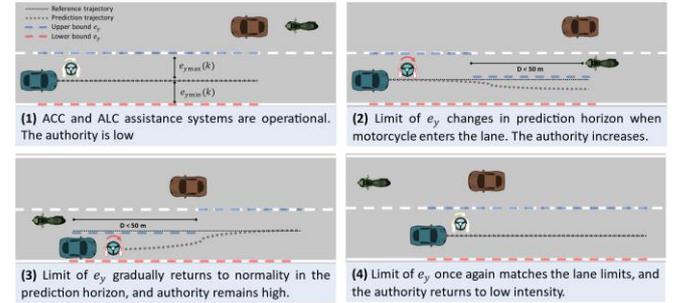

Figure 7. Sequence of scenarios for the study of evasive maneuvers

The lateral error constraint is modified based on the distance $(d(k))$ between the motorcycle and each of the predictions $(k)$ of the ego-vehicle. The lower constraint is set to -1.5 m, and the upper constraint is defined as in (20). The authority $\lambda$ is defined according to (21)

$$e_{y_{max}}(k) = \begin{cases} 1.5 \text{ m}, & d(k) \geq 50 \text{ m} \\ -1.25 \text{ m}, & d(k) < 50 \text{ m} \end{cases} \quad (20)$$

$$\lambda = \begin{cases} 3 \text{ Nm}, & \min(d(k)) \geq 50 \text{ m} \\ 12 \text{ Nm}, & \min(d(k)) < 50 \text{ m} \end{cases} \quad (21)$$

## V. VERIFICATION

To verify the operation of the shared-controller and its associated arbitration system, it is tested for each use case. The NMPC configuration in Table 1 and the NMPC model parameters in Table 2 are used for testing. The ACADOS software [11] is the tool used for solving the NMPC problem, and the automated driving framework is running under the Matlab/Simulink environment.

Table 1. Weights and constraints of the NMPC problem for the use cases of corrective and evasive maneuver

| $[x, u, \Delta u]$ | $W_{cor}$ | $C_{cor}$ | $W_{eva}$ | $C_{eva}$ |
|---|---|---|---|---|
| $X$ | 50 | - | 40 | - |
| $Y$ | 50 | - | 40 | - |
| $\Psi$ | 50 | - | 40 | - |
| $\dot\psi$ | 350 | $\pm 0.4$ rad/s | 300 | $\pm 0.4$ rad/s |
| $e_y$ | - | $[-1.5, 5]$ m | - | $[-1.5, e_{y_{max}}(k)]$ m |
| $T_{mpc}$ | 0.15 | $\pm \lambda$ Nm | 0.2 | $\pm \lambda$ Nm |
| $\dot T_{mpc}$ | 0.25 | $\pm 100$ Nm/s | 0.2 | $\pm 400$ Nm/s |

Table 2. Parameters of the NMPC problem for the use cases of corrective and evasive maneuver

| $m$ | 1650 kg | $J$ | 0.1 kg.m$^2$ | $C_r$ | 118e3 |
|---|---|---|---|---|---|
| $I_z$ | 3234 kg.m$^2$ | $b$ | 0.65 Nm/s | $N$ | 30 |
| $l_f$ | 1.40 m | $k_r$ | 8.77 | $T_s$ | 0.05 s |
| $l_r$ | 1.65 m | $C_f$ | 94e3 | | |

## A. Corrective maneuver

The ego-vehicle is initially traveling at 90 km/h but slows down to 70 km/h when following a slow truck ahead. The driver intends to overtake this truck, but there is a risk of a vehicle appearing in the oncoming lane. Figure 8 shows the response of the shared controller during the corrective maneuver for one of the real drivers in the driver study. From 15 s to 20 s, the driver tries to overtake, but a vehicle appears, decreasing the time-to-collision (TTC). The arbitration system increases the authority $\lambda$, and the stiffness of the controller increases while keeping the control torque below the limit set by $\lambda$. During this intervention, the torque derivative and yaw rate remain below the constraint set in the NMPC formulation.

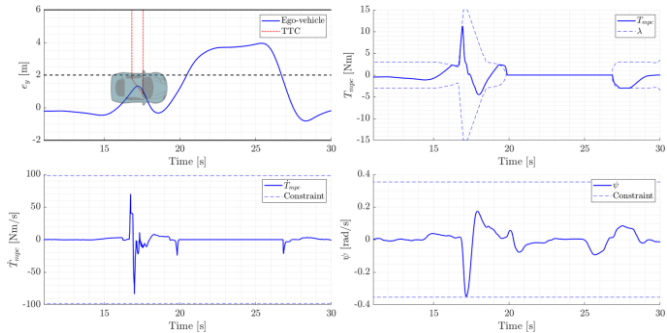

Figure 8. Verification of NMPC for corrective maneuver

## B. Evasive maneuver

The ego-vehicle is traveling at a speed of 105 km/h on a two-way road when a motorcycle suddenly invades its lane. As the distance-to-collision drops below 50 m, the lateral error constraint shifts to guide the vehicle towards the edge of the right lane. Figure 9 illustrates the response of the shared controller for one of the participants of the driver study. The ego-vehicle is directed towards the right with an authority of $\lambda = 12$. The control torque, torque derivative, and yaw rate constraints are all below the NPMC constraint. As a result, the accident was avoided. Once the risk of the maneuver disappeared, the ego-vehicle returned to the center of the lane.

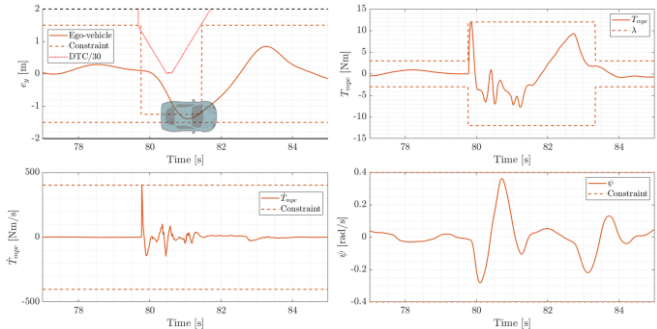

Figure 9. Verification of NMPC for evasive maneuver

## VI. VALIDATION

A study was conducted to validate the shared control approach for both corrective and evasive emergency maneuvers in an immersive dynamic simulator (see Figure 10). Each study involved eight volunteer participants (7 males, 1 female) ranging in age from 25 to 48 years (average = 35.4 years). All participants held a valid driver's license and had at least one year of driving experience (average = 16 years). The majority were regular drivers (covering more than 10,000 km/year, except for one participant with less than 2,000 km/year). Most participants (6 participants) reported some previous interaction with ADAS systems.

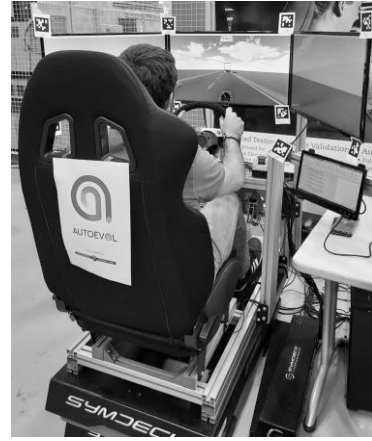

Figure 10. Dynamic driving simulator used for the driver study of critical maneuvers

For objective metrics, a safety assessment [12], [13] is performed for each use case, counting the number of safe and unsafe events. Additionally, statistical analysis of TTC, DTC, and lane deviations is conducted .

For subjective metrics, the User Acceptance Questionnaire by Van der Laan et al. [14] is employed to evaluate the acceptance of the systems by the study participants. Nine items are evaluated using a 5-point Likert scale. The odd-numbered items are averaged to obtain an assessment for the perceived usefulness of the system, while the even-numbered items are averaged to assess satisfaction with the interaction. Furthermore, an ad-hoc questionnaire with four questions on a 9-point Likert scale ranging from "not appropriate at all" to "very appropriate" is used to evaluate the test itself (two questions regarding the sensation of movement in the simulator and the perceived realism of the simulation) and the tested driving support system (one question focused on the appropriateness of the system's intervention, and another one related to the interaction in terms of forces on the steering wheel).

### A. Corrective Maneuver

The shared controller defined in Section 3 is evaluated in conjunction with the decision-making system of Section IV.A. The baseline assumes a vehicle with ACC and ALC operational in the right lane, but without corrective support when initiating the overtaking maneuver. If another vehicle is approaching, the driver must manually execute the correction.

### 1) Objective evaluation

Table 3 compiles the safety metrics for the overtaking corrective maneuvers. The events count shows no crashes and only one accident corresponding to one participant departing the road under SC. As this type of accident leads to stopping the simulation, the total number of events for the SC case is smaller than the number of events on the baseline. Off roads are counted when the vehicle shows a deviation from the center of the lane greater than 1 m.

Table 3. Safety Assessment for corrective maneuver

| KPI | Description | Baseline | Shared control |
|---|---|---|---|
| 1.1 | # Correction events | 46 | 41 |
| 1.2 | # Crashes | 0 | 0 |
| 1.3 | # Near misses | 26 | 13 |
| 1.4 | # Road departures | 0 | 1 |
| 1.5 | # Off-roads | 6 | 4 |

Figure 11 allows comparison of the shared control strategy vs. the baseline. In both conditions there are no crashes with the oncoming vehicle. Yet, the SC brings a 44% reduction in the number of near misses with the oncoming vehicle, while keeping about the same percentage of off-road events. Also, 13.04% off-road excursions for the baseline comes slightly down to 12.2% for the SC yet including one road-departure in which the vehicle completely left the road.

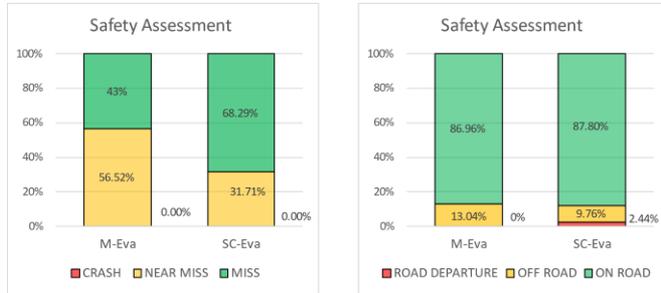
Figure 11. Safety assessment for corrective maneuver

Figure 12 illustrates the results of the TTC and lateral deviation when returning to the lane after a correction.

Regarding the TTC analysis, there is statistical significance ($p = 0.0203 < 0.05$). The SC condition shows a 50% increase compared to the baseline. Furthermore, for the baseline, more than half of the correction events fall below the 0.2 s threshold, while for SC, almost 75% of the events are above 0.2 s.

As for lateral deviation when returning to the lane, there is no significant difference between the two conditions ($p = 0.829$). However, less dispersion is observed in SC, with all data except outliers under 1 m of deviation, which is the limit to be counted as an off-road event.

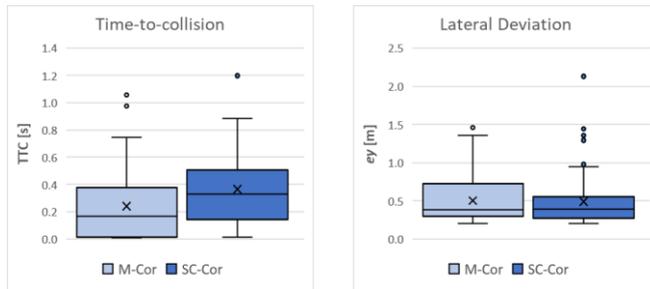
Figure 12. Time-to-collision and lateral deviation results for evasive maneuver.

*2) Subjective evaluation*

Figure 13 provides a direct comparison of user acceptance for corrective overtaking maneuvers supported by shared control versus manually performed maneuvers. Both conditions lie in the positive acceptance quadrant. Regarding satisfaction, there is a slight difference in favor of shared control, although not statistically significant ($p = 0.3064$). However, there is a clear difference in terms of system usefulness, with shared control being perceived as more useful ($p = 0.0067$). This suggests that, according to participant perception, the system's action in correcting maneuvers in dangerous situations is useful, effective, and a good assistant, without compromising good steering wheel interaction, even slightly improving it.

This perception is further supported by the ratings in the specific ad-hoc questions about the support system shown in Figure 14, where participants consider the intervention of the shared control system as highly appropriate ($p = 0.1079$), with only a slight (positive) variation in the assessment of steering wheel force interaction ($p = 0.5983$). The positive evaluation of the system's usefulness and the high rating for corrective intervention correlate with objective measurements demonstrating greater effectiveness in preventing unsafe overtaking, which also leads to good user acceptance.

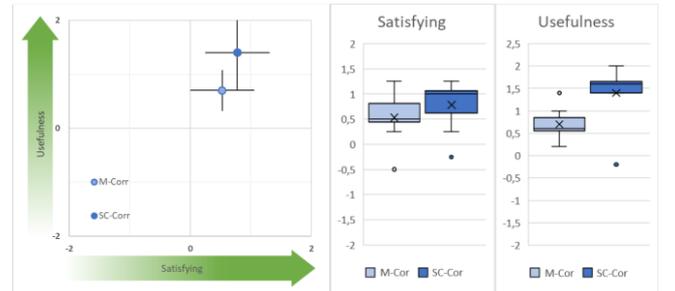
Figure 13. User Acceptance, Corrective Maneuver (Manual vs Shared Control)

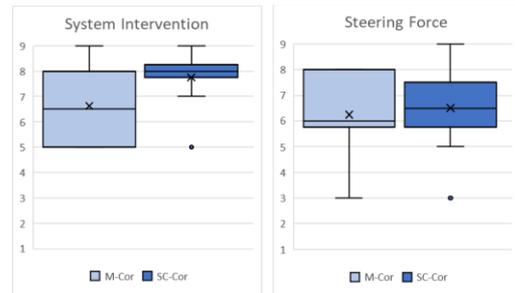
Figure 14. System action perception, Corrective Maneuver (Manual vs Shared Control)

*B. Evasive maneuver*

The shared controller defined in Section 3 is evaluated in conjunction with the decision-making system of Section IV.B. The baseline assumes a vehicle with ACC and no steering support. If a motorcycle invades the lane, the driver must manually execute the evasion.

*1) Objective evaluation*

Table 4 shows the summary of the events recorded in the driver study regarding the evasive maneuver. The smaller number of baseline events is due to one road departure accident that led the end of the simulation before the time of the simulation was completed. Under the baseline mode, 8 accidents were reported, 7 of them directly with the motorcycle and 1 road-departure. Though two crashes were reported under the SC mode, those belong to the same participant, and the other seven shown no crashes.

Table 4. Safety Assessment for evasive maneuver

| KPI | Description | Baseline | Shared control |
|---|---|---|---|
| 2.1 | # Evasion events | 72 | 81 |
| 2.2 | # Crashes | 7 | 2 |
| 2.3 | # Near misses | 28 | 16 |
| 2.4 | # Road departures | 1 | 0 |
| 2.5 | # Off-roads | 41 | 22 |

Figure 15 shows that under the SC, there are 30% more safe events in comparison with the baseline for both misses and on road events. Also, a reduction of almost 50% of the near misses (DTC < 0.2 m) is shown, similarly to the results in the corrective maneuver.

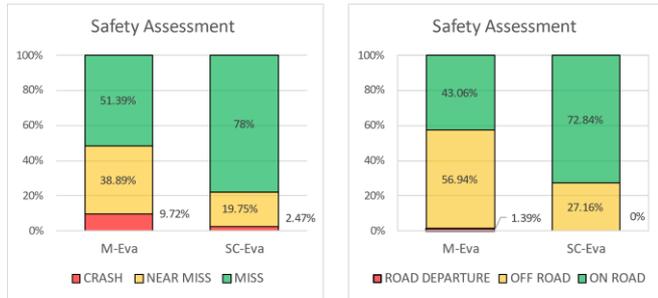

Figure 15. Safety assessment for evasive maneuver

Figure 16 shows the results of the DTC and lateral deviation when returning to the lane after making a correction. In this use case, DTC is a more meaningful measurement than TTC, as the critical nature of the maneuver makes TTC too low and difficult to interpret. The DTC analysis shows no significant difference ($p = 0.1449$), yet, almost 75 % of the evasions under SC are kept above 0.2 m of separation between ego-vehicle and motorcycle, while more than 50 % are below 0.2 m under manual mode.

In terms of lateral deviation, though there is not enough significant difference ($p = 0.0756$), almost 75% of the evasion events under SC are below the limit of lateral error set in the NMPC constraint (1.5 m), while in manual mode more than 50% of the events are above that limit, showing the utility of the constraint in most of the cases.

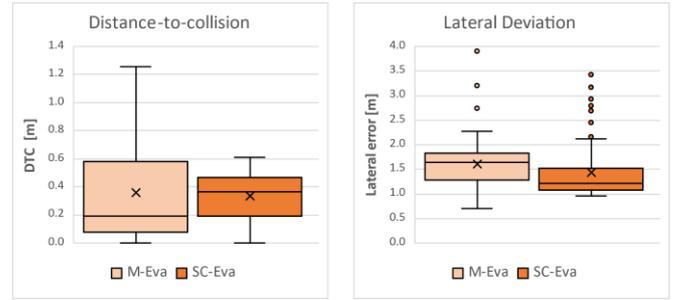

Figure 16. Distance-to-collision and lateral deviation results for evasive maneuver

*2) Subjective evaluation*

Figure 17 allows for a comparison between a manual evasive maneuver and an evasive maneuver with the support of the shared control (SC) system. Both conditions are rated within the positive quadrant. The implementation of the SC support system is perceived as significantly more useful ($p = 0.0229$), in line with the rating of system intervention as very appropriate (Figure 18), and with the significant reduction in recorded accidents. On the other hand, satisfaction ratings are, on average, equal to those in the manual case ($p = 0.8957$), also consistent with the rating of steering wheel force (Figure 18). However, the dispersion in satisfaction ratings reflects the effect of a sudden intervention in an emergency, which some participants describe as unpleasant or annoying but desirable, while others find it pleasant and nice, in addition to being desirable in emergency situations.

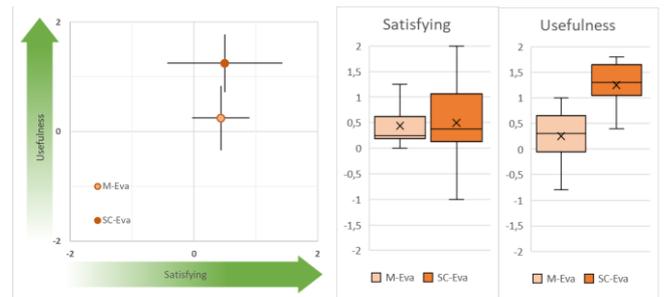

Figure 17. User Acceptance, Evasive Maneuver (Manual vs Shared Control)

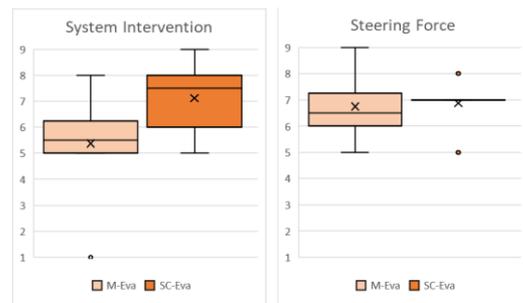

Figure 18. System action perception, Evasive Maneuver (Manual vs Shared Control)

*C. Driving Simulator*

Regarding the evaluation of the simulator, the perception remains consistent in all studied cases, with a positive assessment of the sensation of movement and the realism of the simulation as appropriate (see Figure 19).

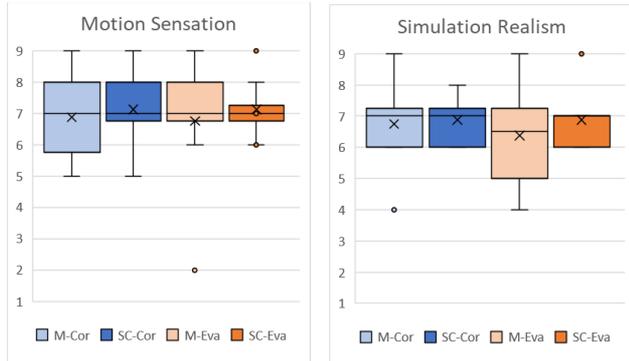

Figure 19. Driving simulation evaluation

## VII. CONCLUSION

Within the Autoev@1 project, the automated driving modular architecture has been revisited to include driver interaction with the automation, through modules for arbitration and for shared controller. Specific decision-making strategies tied to each use case were developed, including a novel strategy to generate the evasive maneuver dynamic path planning through the introduction of soft safety constraints in the MPC. The functionality of the developed modules was verified in terms of the torque and yaw rate constraints, as well as the effectiveness of the soft constraints to generate the evasive maneuver path.

The evasive maneuver validation tests show a significant safety improvement (reduction of accidents and near misses) by avoiding unsafe lane invasions. Which correlates well with the high user acceptance showing an improved usefulness without any loss of satisfaction. On the same lines, the corrective maneuver improves TTC and lateral deviation, thus improving safety (less near misses) together with an improved user acceptance. Hence, the shared control driver assistance strategies developed and tested are effective to improve safety under the considered critical maneuvers and show good user acceptance.

In future works, the implementation of shared control strategies in drive-by-wire systems are expected, to avoid the sometimes unsafe and uncomfortable conflict that is produced in the steering wheel specially for critical maneuvers where the required assistance torque is high.

## ACKNOWLEDGMENT

This work is supported by the Government of the Basque Country by means of AUTOEV@L project (KK-2021/00,123). This work is supported by the EU Commission AWARE2ALL project, which has received funding from Horizon Europe research and innovation programme under grant agreement No 101076868.